# One-Press Control: A Tactile Input Method for Pressure-Sensitive Computer Keyboards

**Staas de Jong**
LIACS
Leiden University
Leiden, The Netherlands
staas@liacs.nl

**Dünya Kirkali**
Media Technology programme
Leiden University
Leiden, The Netherlands
dunyakirkali@gmail.com

**Hanna Schraffenberger**
Media Technology programme
Leiden University
Leiden, The Netherlands
hanna@schraffenberger.de

**Jeroen Jillissen**
Media Technology programme
Leiden University
Leiden, The Netherlands
jeroen@jillissen.com

**Alwin de Rooij**
Media Technology programme
Leiden University
Leiden, The Netherlands
info@alwinderooij.nl

**Arnout Terpstra**
Media Technology programme
Leiden University
Leiden, The Netherlands
dancetrend@gmail.com

## Abstract
This work presents *One-press control,* a tactile input method for pressure-sensitive keyboards based on the detection and classification of pressing movements on the already held-down key. To seamlessly integrate the added control input with existing practices for ordinary computer keyboards, the redefined notion of *virtual modifier keys* is introduced. A number of application examples are given, especially to point out a potential for simplifying existing interactions by replacing modifier key combinations with single key presses. Also, a new class of interaction scenarios employing the technique is proposed, based on an interaction model named *"What You Touch Is What You Get (WYTIWYG)"*. Here, the proposed tactile input method is used to navigate interaction options, get full previews of potential outcomes, and then either commit to one or abort altogether – all in the space of one key depress / release cycle. The results of user testing indicate some remaining implementation issues, as well as that the technique can be learned within about a quarter of an hour of hands-on operating practice time.

## ACM Classification Keywords
H.5.2 Information interfaces and presentation (e.g., HCI): Input devices and strategies.

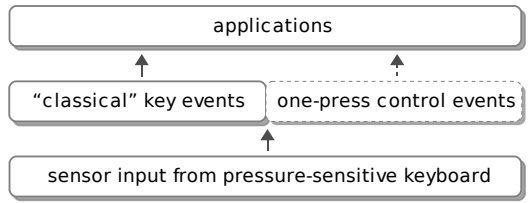

**Figure 1.** The proposed software layer.

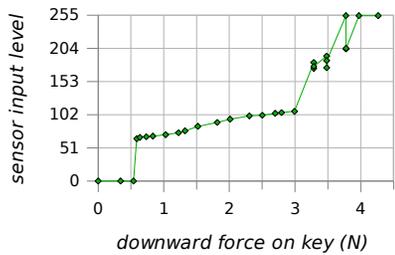

**Figure 2.** Typical response of one key's force sensor.

## Introduction

Recently, Dietz et al. [2] have introduced a pressure-sensitive computer keyboard based on low-cost membrane technology suitable for mass production. The device, in its look and feel exactly like an ordinary keyboard, can independently report the forces applied to those keys that are being depressed. This per-key sensing, and continuous pressure reporting while a key is held down are important additional capabilities when comparing this approach to work in progress reported earlier by Iwasaki et al. [4]. (There, the built-in accelerometer of a laptop was used to estimate typing pressure while striking keys on the laptop's keyboard.) Basically, the newly proposed technology can be said to extend the amplitude resolution of continuous keydown pressure on computer keyboards from 1 bit to more than that.

Naturally, the question then becomes how to best unlock this potential for extra control. Dietz et al. [2] already suggested a number of example techniques: mapping a depressed key's force level directly to some control parameter; use as a low-resolution, multi-touch sensor for gesture recognition; and measuring the force level when striking keys (e.g., to control font size of textual output). Here, we would like to suggest another tactile input method, which we have called *One-press control*.

### One-press control

One-press control enables the control of multiple different events during a single key depress / release cycle. It is implemented as a software layer placed between regular applications and the raw input from the pressure-sensitive keyboard (see Figure 1). Basic functionality can be summarized as follows: "classical" key depress and release events are transparently recreated and passed on. However, if a key is depressed and held down relatively *softly* for some timeout period (similar to that of typematic character repetition), this does *not* lead to a classical key depress event. Instead, the software will now track the relevant force sensor's input in order to look for subsequent and possibly repeated *pressing movements* on the already held-down key. Figure 2 shows a typical response of one force sensor, indicating a usable input range between 0.6 and 3.0 Newton. This overlaps with the 0 - 3 Newton range from Mizobuchi et al. [5], given there as a reasonable choice (in terms of performance and comfort) for the pen-based control of user interfaces.

**Figure 3.** Force peak extraction after an initial soft depress. (Sensor data sampled after some conditioning.)

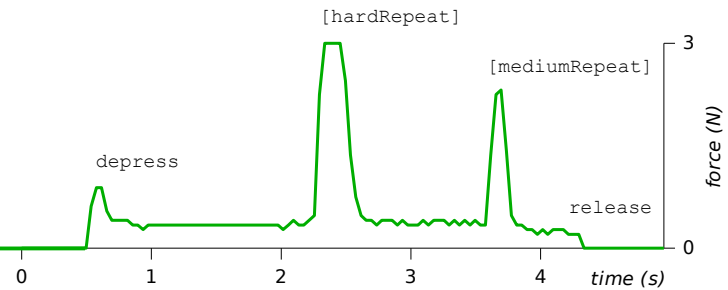

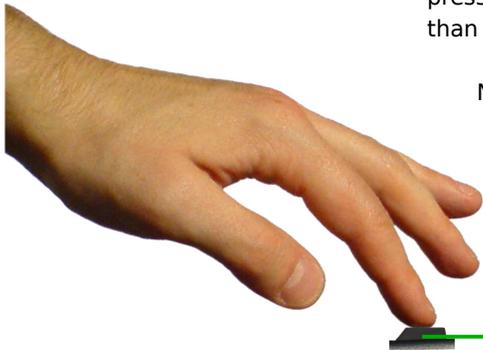

Using the force sensor's input, pressing movements are then detected as discrete events and labeled `[mediumRepeat]` or `[hardRepeat]`, according to their intensity (see Figure 3). This may be compared to other research where pressing events of varying amplitude have also been extracted from tactile input employing fixed flat contact surfaces. One example of this is the "Quick release" pressure technique described by Ramos et al. [6] and used (in the context of touchscreen mobile devices) by Brewster and Hughes [1]. An important difference, however, is that in the current situation the finger is not necessarily lifted from the surface after a pressing event, and detection is not triggered by this. Another example of prior research is that by Rekimoto and Schwesig [8], where finger pressure input on a touchpad initiates and terminates layered depress states as it crosses force thresholds. In the current approach, force thresholds are also used to classify amplitude, but pressing events are now regarded as atomic in nature, and detected based on the first-order derivative of force input.

After detection, the label assigned to an input force peak is passed on to applications by attaching it to the source key in question as a *virtual modifier key*. For example, in addition to a classical event like `[alt][F1]`, an application may now also see and respond to a `[mediumRepeat]` or `[hardRepeat][F1]`.

*Some potential advantages*
The approach described so far presents a number of potential advantages:

• Augment, rather than replace: keyboard interaction is extended while leaving traditional mechanisms in place.

• The required skill of making pressing movements may be more familiar to keyboard users than, say, having to control steady force levels to provide input.

• While more control becomes directly accessible, "bailing out" of any new-type interactions can be made as easy and intuitive as a timely key release.

• Using virtual modifier keys, developers can plug new-type key events right into existing applications. For example, by using a `[hardRepeat][del]` just like `[shift][del]`, to permanently delete a file; or by using a `[hardRepeat][a]` just like `[shift][a]`, to type "A" instead of "a".

• Awkward modifier key combinations may be replaced by single key presses, simplifying interaction. For example, by using a `[hardRepeat][F4]` instead of `[alt][F4]` to close a program window; or by using a `[mediumRepeat][tab]` instead of `[alt][tab]` to switch to the next window.

**More application examples: WYTIWYG**
In order to further illustrate and assess the potential applicability of one-press control, a number of demos implementing example scenarios of its use have been created. Behind each scenario is the overall goal of navigating interaction options in a user interface. This is done following an interaction model based on two design principles:

• For tactile input: Let the user navigate interaction options and express confidence in them through tactile force.

• For visual output: Give previews of possible interaction outcomes, in a way matching actual output as closely as possible.

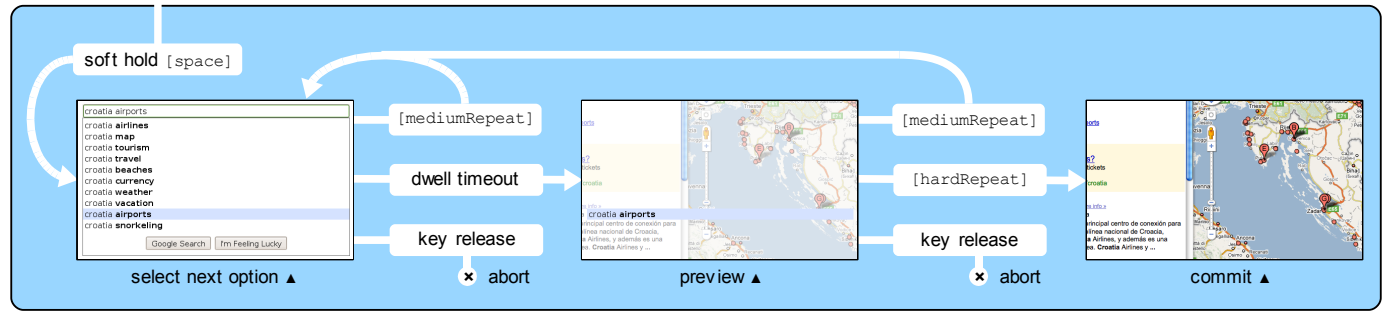

**Figure 4.** WYTIWYG example scenario: one-press control of Google Suggest using the `[space]` key.

Other implemented variations, increasing in exploratory sophistication:

*"I could be..."*

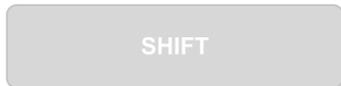

*"...pressing this key."*
("Blind" typing for everyone.)

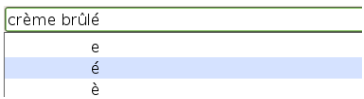

*"...typing this word (correctly)."*

[Croatian] Hrvatska "izlet brodom" Zadar sno
[French] Croatie "voyage en bateau" Zadar "p
[Chinese] 克罗地亚 乘船游览 扎达尔 浮潜
[Dutch] Kroatië boottocht Zadar snorkelen
[Italian] Croazia "gita in battello" Zadar snorke

*"...using these search term translations."*

Many more examples of this type of interaction are imaginable.

We named this approach "*What You Touch Is What You Get* (*WYTIWYG*)", analogous to the well-known "What You See Is What You Get (WYSIWYG)" paradigm.

The approach is similar in nature to the "previewable user interfaces" presented in Rekimoto et al. [7]. There, in the instances where preview of an option is activated by *touching* a key, and commitment by *pressing* it, tactile force can also be said to express confidence. Furthermore, both approaches aim to replace trial-and-error interactions (based on the execution of do-undo actions) with more explorative interactions, where there is room for doubt and variable levels of confidence in the interaction options available. Still, an important difference in the current approach is its explicit emphasis on providing a preview which, while still flagged as such, in fact allows the user to inspect it at the same level of detail as actual output. This is intended to maximize a preview's usefulness in informing subsequent actions in the exploratory interaction process.

*A typical scenario*
Since typing is the main application of computer keyboards, implemented example scenarios have focused on this aspect of keyboard input. In a typical scenario in the context of web searching, illustrated in Figure 4, a hypothetically extended version of Google Suggest [3] is controlled via the `[space]` key:

• While composing a search query, the user softly presses and holds `[space]`. After a short timeout, this activates entry into a dropdown menu containing possibly relevant alternatives for text input.

• Now, repeated `[mediumRepeat]` presses cycle through the available items (also expressing a tentative confidence in them). Dwelling on an option activates a preview of the associated search results. The preview is distinguished from actual output only by having a reduced visual contrast, and by having the selected menu item still visible inside it.

• At any point, the user may "bail out" from the presented alternatives, by simply releasing `[space]` again (expressing zero confidence). Alternatively, a firm final `[hardRepeat]` (expressing maximum confidence) commits to the currently selected option.

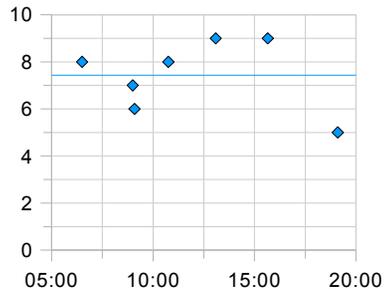

**Figure 5.** Test subjects' performance in terms of operating practice time (horizontal, min:sec) and perfect task executions (vertical).

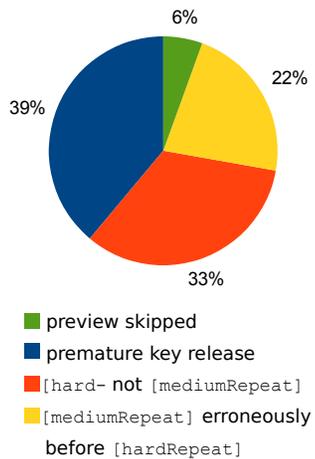

**Figure 6.** Classification of failed task executions.

## Evaluation

In order to assess the learnability of the one-press control tactile technique and identify weak spots in its current implementation, an experimental evaluation was performed. This was done using the example scenario described in the previous section, as it utilizes all aspects of the technique and places them in a realistic user interface context.

7 volunteer test subjects took part in the experiment: 3 females and 4 males, aged between 23 and 40 years old. All were new to the keyboard and input technique. None reported having any visual, manual or other impairments relevant to everyday computer use. All rated themselves between "somewhat" and "very" familiar with computer keyboard use.

*Procedure*
Each test subject would sit at a table with a Microsoft experimental pressure-sensitive keyboard placed within easy reach. A Macbook laptop screen was used for visual output, showing a simulated browser window. First, a short introduction to the tactile input method would be given. Then, subjects familiarized themselves with the task to be evaluated, in 4 cumulative practice stages: (1) Performing [mediumRepeat] presses on one specific key to navigate to arbitrary menu items (10 items total). (2) Holding the key steady when on an option, until the associated preview would appear. (3) Performing [hardRepeat] presses after such preview activations, to commit to the related search result. (4) Doing this for a fixed target: menu item #8.

Before each practice stage, the subject would be shown how to handle the key, and the correct visual feedback to expect. When making mistakes while practicing with the key, the test subject would receive verbal feedback indicating what was going wrong and how it could be corrected. This would continue until the test subject indicated being comfortable with the technique so far. Test subject key operating practice time was explicitly delimited, and recorded at the end of each stage.

After completing practice stage 4, test subjects were asked to execute its task 10 times in a row to the best of their ability. The first 10 subsequent attempts were then logged and automatically classified as successes or failures. Any key depress / release cycle not navigating to the correct option, activating its preview, and committing by a [hardRepeat] would be classified as a failure. (This included, among many other things, quick aborted keypresses.)

*Results*
The main results regarding learnability can be seen in Figure 5. As is plotted, the total operating practice times of participants ranged from 6:30 to 19:06 min:sec. Final task scores ranged from 5 to 9, with participants scoring 7.4 perfect executions on average. All subjects rated final task easiness between "neutral" and "easy", except for the subject having both the lowest score and the longest practice time, who rated it "hard". It should be noted that this participant did log 9 perfect executions in a row. However, only 4 of these were still part of the first 10 attempts counted.

Analyzing the trial logs, it was possible to classify all failed attempts (18 out of 70) into 4 categories, which are shown in Figure 6. The three main causes turned out to be two different types of mix-ups between [hardRepeat] and [mediumRepeat] presses, and unintended key releases. The latter seemed the dominant issue, especially since 5 test subjects noted avoiding premature key releases as one of the most difficult aspects of the final task.

## Conclusion and future work

We have introduced *One-press control*, a tactile input method for pressure-sensitive keyboards based on the detection and classification of pressing movements on the already held-down key. As a mechanism to seamlessly integrate this with existing practices for ordinary computer keyboards, we proposed the redefined notion of *virtual modifier keys*. We then gave some concrete examples of how the technique could be used in a range of existing user interface situations, especially pointing out a potential to simplify existing interactions by replacing modifier key combinations with single key presses. Also, we described how the technique could be applied to a new class of interaction scenarios, based on an interaction model we named *"What You Touch Is What You Get"*. Here, one-press control is used to navigate interaction options, get full previews of potential outcomes, and then either commit to one or abort altogether – all in the space of one key depress / release cycle.

Using one such scenario, user testing of the tactile input method was performed, in order to assess its learnability and identify weak spots in the current implementation. The results indicated that it is possible for people to learn the technique within about a quarter of an hour of hands-on operating practice time, and then execute it with a reasonable degree of perfection. Still, a number of issues were identified as well, most prominently the occurrence of unintended key releases. Before performing any future evaluations comparing the approach to established input techniques, it should be ascertained to what extent adjusting the force event detection algorithm can help solve this issue.

Finally, one currently inherent disadvantage of the approach is that the layered pressing movements lack specific tactile feedback confirming their execution, e.g. such as that implemented using piezo actuators in Rekimoto et al. [8]. Perhaps a future version of the keyboard technology, or adding other forms of feedback (such as audio) could help to address this.

**Acknowledgments**
We gratefully acknowledge Paul Dietz and the Microsoft Applied Sciences Group for providing us with keyboard prototypes. Thanks also to Bas Haring and the Media Technology programme at LIACS, Leiden University for supporting this project. The first author would like to thank his co-authors for their sustained efforts in participating in this project. Thanks to all test subjects for kindly volunteering.